\documentclass[sigconf]{acmart}
\AtBeginDocument{%
  \providecommand\BibTeX{{%
    \normalfont B\kern-0.5em{\scshape i\kern-0.25em b}\kern-0.8em\TeX}}}

\acmConference[KDD '21]{KDD '21: The Second International MIS2 Workshop: 
Misinformation and Misbehavior Mining on the Web}{Aug 15, 2021}{Virtual}
 \acmBooktitle{KDD '21: The Second International MIS2 Workshop: Misinformation and Misbehavior Mining on the Web,
   Aug 15, 2021, Virtual}
\graphicspath{{img/}}
\usepackage{CJKutf8}
\begin{document}
\title{
Dataset of Propaganda Techniques of the State-Sponsored Information Operation of the People's Republic of China}

\author{Rong-Ching Chang}
\email{g08350003@thu.edu.tw}
\affiliation{%
  \institution{Tunghai University}
  \country{Taiwan}
}
\author{Chun-Ming Lai}
\email{cmlai@thu.edu.tw}
\affiliation{%
  \institution{Tunghai University}
  \country{Taiwan}
}

\author{Kai-Lai Chang}
\email{g08350009@thu.edu.tw}
\affiliation{%
  \institution{Tunghai University}
  \country{Taiwan}
}

\author{Chu-Hsing Lin}
\email{chlin@thu.edu.tw}
\affiliation{%
  \institution{Tunghai University}
  \country{Taiwan}
}
% \author{}
% \email{}

\renewcommand{\shortauthors}{Chang, Lai, et al.}

\begin{abstract}

The digital media, identified as computational propaganda provides a pathway for propaganda to expand its reach without limit. State-backed propaganda aims to shape the audiences’ cognition toward entities in favor of a certain political party or authority. Furthermore, it has become part of modern information warfare used in order to gain an advantage over opponents.

Most of the current studies focus on using machine learning, quantitative, and qualitative methods to distinguish if a certain piece of information on social media is propaganda. Mainly conducted on English content, but very little research addresses Chinese Mandarin content. From propaganda detection, we want to go one step further to providing more fine-grained information on propaganda techniques that are applied.

In this research, we aim to bridge the information gap by providing a multi-labeled propaganda techniques dataset in Mandarin based on a state-backed information operation dataset provided by Twitter.
In addition to presenting the dataset, we apply a multi-label text classification using fine-tuned BERT. Potentially this could help future research in detecting state-backed propaganda online especially in a cross-lingual context and cross platforms identity consolidation.

\end{abstract}





\maketitle

\section{Introduction}

Propaganda has the purpose of framing and influencing opinions. With the rise of the internet and social media, propaganda has adopted a powerful tool for its unlimited reach, as well as multiple forms of content that can further drive engagement online and offline without disclosing the writers’ identity. Computational propaganda is defined as propaganda being created or distributed using computational or technical means \cite{bolsover2017computational}. Exploiting social media is considered as one of the low-cost and high-impact techniques in information warfare, driving and manipulating human behavior with various psychological manipulations \cite{10.2307/26481910}. How information is conveyed is by using propaganda techniques. Propaganda techniques are not only used for political content, but also for marketing, and religious content for persuasion purpose. Propaganda techniques, commonly used in disinformation and misinformation, are the way that propaganda is conveyed \cite{da2019fine}, such detection allows <requires?> for more fine-grained analysis and detection, not only distinguishing if it is propaganda, but characterizing where it might come from. The propaganda activity launched by foreign adversaries could be particularly concerning to a country as the usual goal may include steering discord, spreading fear, influencing beliefs and behaviors, diminishing trust, and threatening the stability of a country  \cite{10.2307/26481910}. Various state-backed official and unofficial departments, organizations, and agencies were established to address information warfare include the Internet Research Agency of Russia \cite{diresta2019tactics}, 50 Cent Party \cite{king2017chinese} \cite{han2015manufacturing} of Chinese Communist Party (CCP) and the Public Opinion Brigades of the Communist Party of Vietnam \cite{bradshaw2017troops}.

Most of the recent work has been focused on propaganda detection, in other word, identifying if the information is propaganda or not. This has been done using various methods such as qualitative analysis, quantitative analysis \cite{beskow2020characterization}, and machine learning \cite{wickramarathna2020framework} \cite{rcc}. The main features for this detection task could be divided into two parts, content-driven, and network-driven. 
Some of the current propaganda text corpora open data sets on document levels include \citet{rashkin2017truth} which labeled texts into trusted, satire, hoax, and propaganda on news. \citet{barron2019proppy} further increased the corpus \cite{rashkin2017truth} with more propaganda articles and metadata of the articles. The currently available fine-grained propaganda technique dataset is the one presented by \citet{da2019fine}. From news articles, they labeled 18 propaganda techniques on a word-by-word sentence level, so that the position of where the propaganda technique was applied from start to end was being documented. All of the mentioned data sets are in English. \citet{baisa2019benchmark} released a propaganda technique dataset for Czech based on newspaper. 
 Another open data source is Twitter, a popular social media platform, the dataset discloses state-linked information operations that took place on their platform. 
 However, the Twitter dataset is not labeled with propaganda techniques but the Twitter account metadata and media information only. The previously labeled propaganda technique in news article texts could be quite different linguistically compared to texts on social media. Tweets, messages posted on Twitter, tend to be more casual with slang and emoji. They are also shorter as the platform has a text length limit.
In the literature \citet{da2020survey} who conducted a survey of computational propaganda, mentioned that there is limited propaganda detection research based on text features due to the lack of annotated data sets. Yet we think text content is an important feature for performing cross-platform detection, in user-identity linking, and in information origin tracing. Since the network feature may differ from platform to platform, text content is more consistent in that regard. To our knowledge, there is no existing propaganda technique dataset for Mandarin  Chinese.   

To address such a gap, we present our dataset\footnote{Dataset will be released on https://github.com/annabechang} that focuses on propaganda techniques in Mandarin based on a state-linked information operations dataset from the PRC released by Twitter in July 2019. The dataset consists of multi-label propaganda techniques of the sampled tweets. Additionally, we employed a fine-tuned BERT model for the multi-label classification task.

\section{Propaganda Techniques}

Below we explained a list of selected propaganda techniques we have considered based on various studies \cite{da2019fine} \cite{baisa2019benchmark} \cite{enwiki:1020793767}. Using the same assumption as \cite{da2019fine}, we labeled our data based on the linguistic and language use that can be judged directly without retrieving extra information. The propaganda techniques we considered are as follows: 

\begin{enumerate}

  \item Presenting Irrelevant Data
  
  Also called Red Herring. Introducing irrelevant information or issues to an argument. 
  
  \item Misrepresentation of Someone's Position (Straw Man)
  
  Substituting one's opinion with a distorted version rather than the original one.    
  
  \item Whataboutism
  
  Defaming the opponents with hypocrisy.
  
  \item Oversimplification \\
  Overly generalizing information or the complexity of the certain issues to favor a party. 
  
  \item Obfuscation, intentional vagueness, confusion\\
  Purposefully being vague with the intention for the audience to develop false recognition toward the subject. 
  
  \item Appeal to authority\\
  Supporting the opinion or clam unconditionally as long as it comes from the government or an expert.
  
  \item Black-and-white Fallacy\\
  Presenting only two opposite possibilities, one favoring a certain party and one presented by the opponent. 
  
  \item Stereotyping, name-calling, labeling\\
  Labeling the target with the intention of arousing prejudices or making an association with stereotypes. 
  
  \item Loaded Language\\
  Using emotional words to influence audience opinions. 
  
  \item Exaggeration or Minimisation\\
  Overly amplifying or reducing the importance of something. 
  
  \item Flag-waving\\
  Justifying or presenting as a nation or group or idea. In our case, we also consider Flag-waving when one person is presented as their opinion represents the entire nation or group.
  
  \item Doubt\\
  Questioning or steering uncertainty or trust toward something, an entity, or a group.
  
  \item Appeal to fear or prejudice \\
  Spreading a sense of anxiety, fear, or panic toward the audience or entity.
  
  \item Slogans \\
  A brief sentence that includes labeling, stereotyping or certain cognitive belief.
  
  \item Thought-terminating cliché\\
  Using simple and generic sentences to discourage detail in discussions.

  \item Bandwagon\\
  Persuading the audience to align with the bigger crowd who appear to have an advantage or better situation, or implying a certain entity will lose or have a worse situation.  
  
  \item Guilt by association or Reductio ad Hitlerum\\
  Associating an opponent or target with the usually disliked object or group.
  
  \item Repetition\\
  Repeating the same message or idea several times. 
\end{enumerate}

Additional to the usual propaganda techniques, we also introduce the following that have been seen in the dataset:

\begin{enumerate}
  \item Neutral Political \\
  This includes the international political news that's being written objectively. 
  
  \item Non-Political\\
    This includes the non-political related content, which could be written with a neutral or angry, or happy tone. 

  \item Meme humor\\
  This is the content that used sarcastic humor toward an entity.
\end{enumerate}

\section{Data}

Twitter disclosed 936 accounts with identified state-backed information operations from the People’s Republic of China (PRC) government departments, agencies, and party-supported organizations. The dataset was divided into two batches that consist of 744 accounts and 196 accounts separately on Twitter's information operations disclosure. In our study, we sampled tweets from a batch of 744 accounts. The available data disclosed containing account metadata (account created time, a user profile description, user-reported location, etc), tweet metadata (tweet created time, tweet text, tweet language, tweet hashtags, etc), and shared media content (images and videos). In our study, we only focus on the tweet metadata.  

The total number of tweets sent by the 744 accounts is $ 1,898,108$. We first filter it by language, and duplicates were dropped. The total number of tweets in Chinese contained in the dataset is $ 74,277$, we randomly selected $ 9,950$ tweets out of that number for labeling. 

\begin{figure}[ht]
  \centering
  \includegraphics[width=\linewidth]{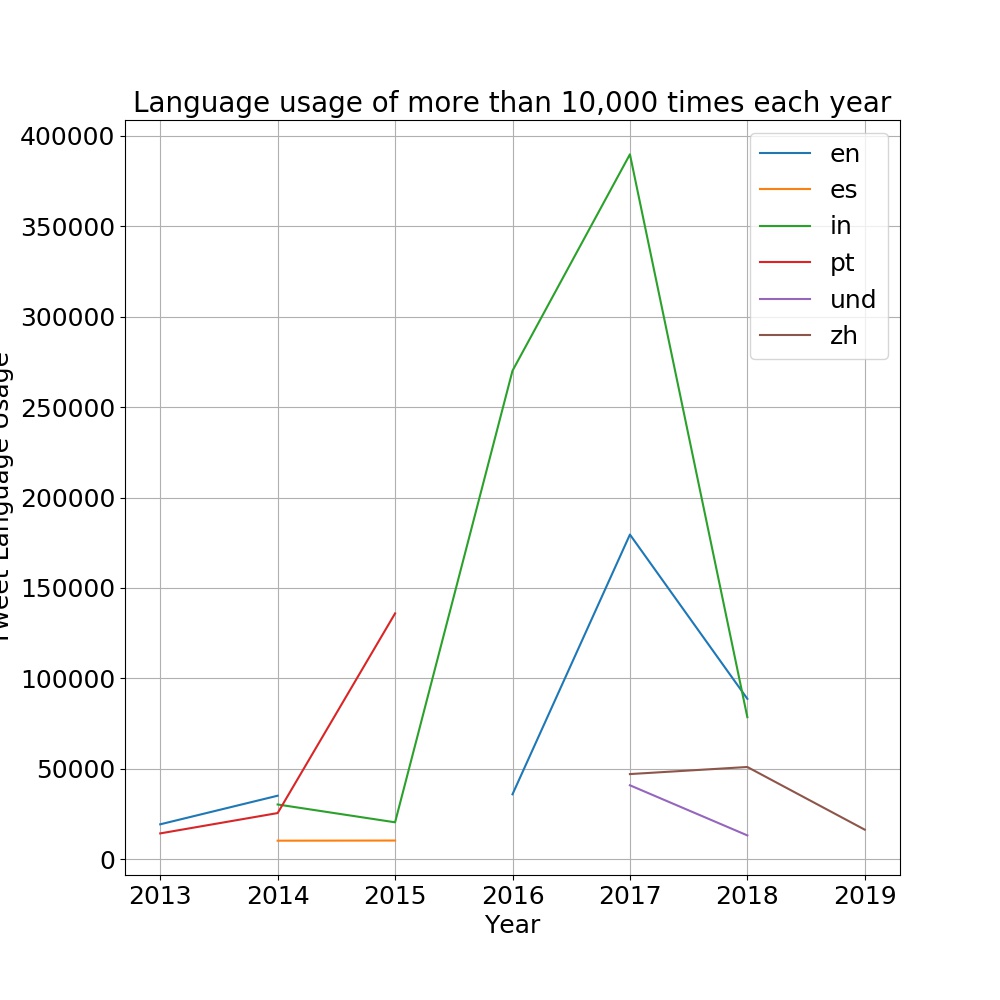}
  \caption{Total language usage in the data set}
  \label{fig:total_lang}
\end{figure}

\citet{uren2019tweeting} conducted a detailed quantitative and qualitative analysis on these accounts, and suggested that this cluster could be re-purposed spam accounts as they altered the used language through different periods of time. These findings are aligned with ours. Figure \ref{fig:total_lang} shows the top 15 tweet language usage out of 50 total used languages. The top 5 languages used in this cluster of accounts are  Indonesian (in), English (en), Portuguese (pt), Chinese (zh), and Tagalog (tl). 

\begin{figure}[ht]
  \centering
  \includegraphics[width=\linewidth]{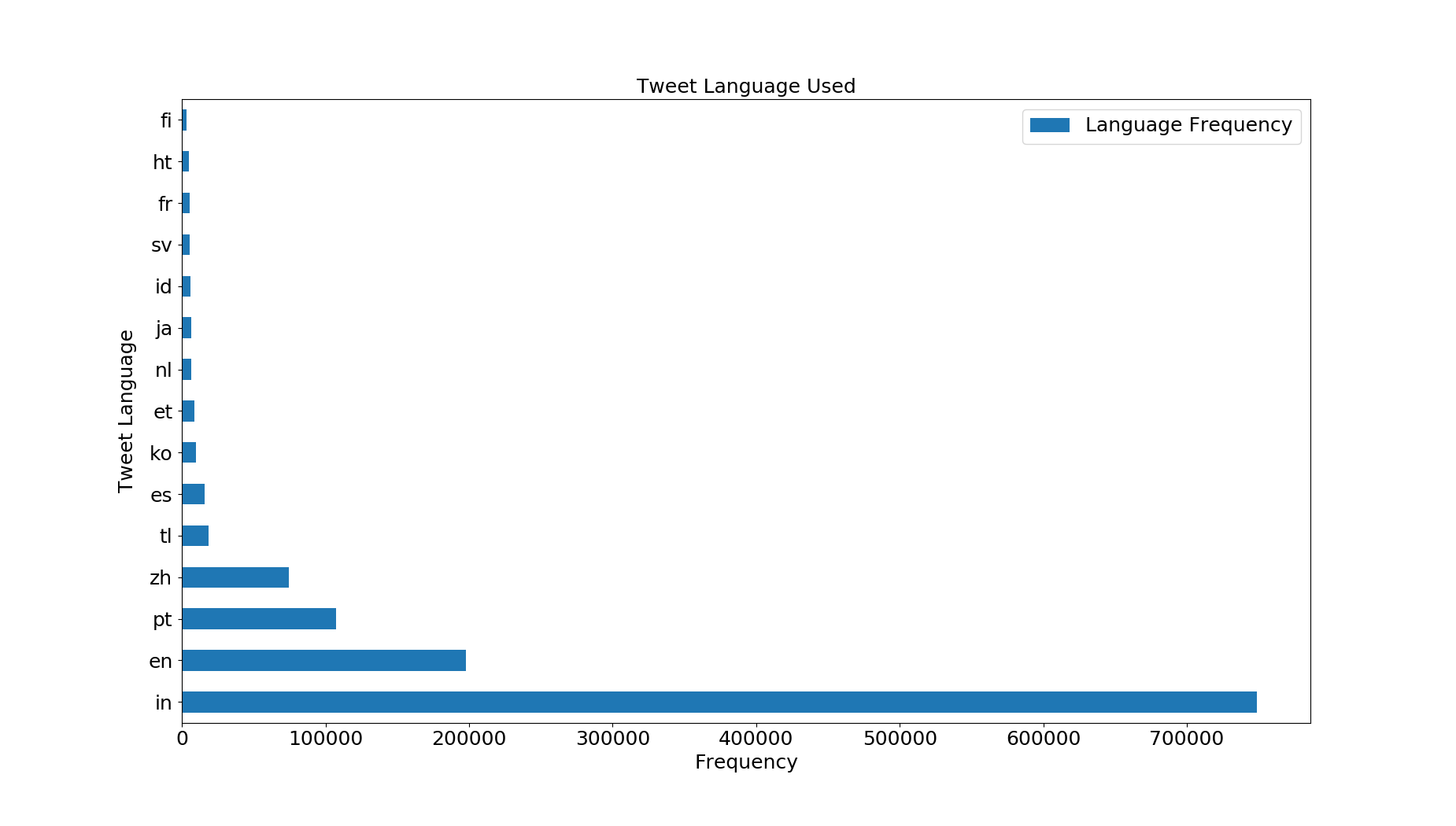}
  \caption{Language usage of more than 10,000 times each year}
  \label{fig:lang_used}
\end{figure}

In Figure \ref{fig:lang_used} we plot the language used more than 10,000 times each year and we can see that Chinese was only used by this cluster of accounts after 2017. The primary used language appears to have been clear cut in different years, which indicate that this cluster could be spam accounts that were created and used by entities with different backgrounds and purposes at different time period. 

Manual annotation was done by two annotators separately on different portions of the dataset, this was designed intentionally to insure the alliance of opinions in the dataset \cite{gordon2021disagreement}. This design will increase the annotator consistency, reduce noise and have better model performance. In the annotation process, we iterate the process of reviewing, data labeling and documenting political entities being named in the sentence. We built pre-defined labels according to the political entities mentioned, and human annotators update from the rule based labels, adding more entity keywords \footnote{The list of entity keywords will be released on the same website}, and updating the further unseen data labels. The keywords are show in Table \ref{tab:key}, they can be divided into four categories: exiled or anti-government Chinese, Hong Kong protest, Taiwan independence, International Geo-Political related topics. The usage of keyword is not to be exact but assisting the human annotators.

\begin{table}[ht]
  \caption{Aggregation of keyword mentioned count}
  \label{tab:key}
  \begin{tabular}{cc}
    \toprule
    Keyword Category & Count \\
    \midrule
    Exiled or anti-government Chinese & 5,406 \\
    Hong Kong protest & 209 \\
    International Geo-Political & 1,718 \\
    Taiwan independence & 2 \\
  \bottomrule
\end{tabular}
\end{table}

The propaganda techniques are only labeled on the political-related content, there could be non-political content using propaganda techniques but this is not labeled as it was not our focus. Such content will be labeled as a non-political class.   

In total, we have 21 different propaganda techniques, we showed a label statistic in Table \ref{tab:freq}. This is an imbalanced dataset, as the most frequently used label is the non-political content that was used for $ 6,117$ times.  Loaded Language was used the most at $ 2,609$ times, followed by Whataboutism $ 2,509$ and Name-Calling $ 2,313$ are the most used propaganda techniques on political-related content. A few techniques occurred rarely, especially the Thought-terminating cliché was not used. We suspect that this is due to the nature of spam accounts. That is, building a relationship with other accounts was not their primary goal. Thought-terminating clichés might be used more in the circumstances where building a  relationship with other accounts is one of the target goals. An example of how the dataset was formatted can be seen in Table \ref{tab:smp}, the tweets were translated for the purpose of display. 

\begin{table}[ht]
  \caption{Data set label statistics}
  \label{tab:freq}
  \begin{tabular}{ccc}
    \toprule
    Symbo & Propaganda Techniques & Frequency\\
    \midrule
    1 & Presenting Irrelevant Data & 13\\
    2 & Straw Man & 2\\
    3 & Whataboutism & 2,509\\
    4 & Oversimplification & 37\\
    5& Obfuscation& 12\\
    6& Appeal to authority& 50\\
    7& Black-and-white & 265\\
    8& Name Calling& 2,313\\
    9& Loaded Language& 2,609\\
    10& Exaggeration or Minimisation& 114\\
    11& Flag-waving& 81\\
    12& Doubt& 147\\
    13& Appeal to fear or prejudice & 141\\
    14& Slogans& 37\\
    15& Thought-terminating cliché& 0\\
    16& Bandwagon& 64\\
    17& Reductio ad Hitlerum& 83\\
    18& Repetition& 60\\
    19& Neutral Political & 915\\
    20& Non-Political& 6,117\\
    21& Meme humor& 5\\
    
  \bottomrule
\end{tabular}
\end{table}

\begin{table}[ht]
  \caption{Data set sample display}
  \label{tab:smp}
  \begin{tabular}{|p{1.5cm}|p{4cm}|p{1.5cm}|}
    \toprule
    Tweetid & Translated Tweet & Propaganda Techniques\\
    \midrule 
    \shortstack[r]{990189929\\836699648} & The truth and hypocrisy under  the false democratic face of Guo Wengui, the clown jumping beam, is now undoubtedly exposed! & 3,8,9\\
    \hline 
    \shortstack[r]{114879827\\6281364480} & We must severely punish the rioters and return Hong Kong to peaceful & 8,9,13,14\\
  \bottomrule
\end{tabular}
\end{table}

\section{Multi-label Propaganda Technique Classification}
In this section, we describe our methodology in designing andfine- tuning, and provide the result of our BERT-based multi-label classification result.

Bidirectional Encoder Representations from Transformers \(BERT\) \cite{devlin-etal-2019-bert} a language representation model has delivered state-of-the-art results in several NLP tasks. In our case, the research problem in our case is a multi-label task where given one sentence, there are one to multiple labels that could apply. 

We used the bert-base-chinese pre-trained model provided by Huggingface \cite{wolf-etal-2020-transformers} for both tokenization and pre-training the model. The bert-base-chinese pre-trained model is trained based on both simplified Chinese and traditional Chinese \cite{cui2019pre}, which fits our use case. In our model design, we used a BERT model followed by a dropout and linear layer for regularization and classification purposes. We have 21 different labels defined in our propaganda technique labels with 1 of them without occurrence. We set the number of dimensions for the linear layer to 20. The output of the linear layer is what we used to determine the accuracy of the models. 

The max input length was set to 100 with a training batch size of 2 and a validation batch size of 2 using the data loader from Pytorch \cite{NEURIPS2019_9015}. We chose to use BCEWithLogitsLoss, which combines a Sigmoid layer and BCELoss, from Pytorch  \cite{NEURIPS2019_9015} as our loss function. Adam \cite{1412.6980} was used as an optimizer. We ran it for 2 epochs with a learning rate equal to $ 1e-05$. We trained on a Linux machine with GeForce RTX 2070 GPU, and 16 Intel(R) Core(TM) i9-9900K CPU.

\section{Evaluation}
The training and testing size was set to 80\% and 20\% respectively. The results are shown in the Table \ref{tab:res}. We only trained it for 2 epochs yet we saw the loss decreased drastically from $ 0.71102$ to $ 0.05953$. In the experiment, we trained for more than 2 epochs; however, the accuracy did not improve. Thus 2 epochs appear to be optimal in our experiment. The evaluation metrics used were accuracy, micro-averaged F1-score, and macro-averaged F1-score. Micro-averaged F1-score aggregate all the samples to compute the average metric of the True Positives our of the Predicted Positives. Macro-averaged F1-score aggregated each class and compute the metrics based on each class. In our case, our accuracy is $ 0.80352$ with micro-averaged F1-score of $ 0.85431$ and macro-averaged F1-score of $ 0.20803$. This indicates that our model performed well in predicting overall samples, however the performance on each label varied a lot. This is expected as our dataset is skewed, some labels have many data while a few labels have very little data labeled in the dataset. 

\begin{table}[ht]
  \caption{Classification results}
  \label{tab:res}
  \begin{tabular}{cc}
    \toprule
    Measurement Name & Performance \\
    \midrule
    Loss : Epoch 0 & 0.71102 \\
    Loss : Epoch 1 & 0.05953 \\
    Accuracy& 0.80352 \\
    F1 Score (Micro) & 0.85431 \\
    F1 Score (Macro) & 0.20803 \\
  \bottomrule
\end{tabular}
\end{table}

% \section{Related work}

Two main activity directions of the dataset were to target opponents of the CPC, such as exiled Chinese, human rights lawyers, relevant personnel and to vilify the protesters against the national security law in Hong Kong. This finding was aligned with what was found in \cite{uren2019tweeting} \cite{bolsover2017computational}, where the spam accounts flooded content in Mandarin with the purpose of dominating search results on Twitter when it comes to certain topics. By doing so the propaganda operators wanted the search results to be skewed toward a perspective that favored the CCP and eschewed the certain community.

\section{Discussion}
In this paper, we presented the first propaganda technique dataset of state-backed information operation accounts from PRC for Mandarin based on dataset released by Twitter. We applied 21 propaganda techniques and we annotated a total of $ 9,950$ sentences under a multi-label setting. Machine learning models driven propaganda research can be particularly benefited by our data set. As we labeled political content with propaganda techniques, while giving non political item a label. Our dataset can be used to train classifier for political and non-political in Mandarin as well.

Upon the organization structure of PRC, different departments and agencies may lunch online operations targeting the same or different groups of audiences, with different linguistic characteristics. Thus, this data set's linguistic feature or the propaganda techniques may not apply to all. 

\section{Conclusion}

We presented a new dataset on propaganda techniques from the state-backed information operation accounts from PRC in Mandarin. We trained a fine-tuned BERT model to perform multi-label classification on our dataset. In the times where information on social media are part of information warfare strategics. Our dataset could be beneficial in the propaganda, political research and beyond.

By considering the country, political party, or authority as an entity, we could initially view state-backed propaganda on different topics as a stance detection of texts from such an entity. And propaganda techniques could be viewed as a writing style feature. This could help future research in clustering and identifying how likely it is that the information is coming from the same entity or agency.

One state could launch several propaganda texts that have a similar stance or opinion in different countries with different languages. Thus we hope to see our dataset inspire or provide useful information on multilingual or cross platform state-back propaganda research, using the propaganda techniques as the universal features across languages. 

\bibliographystyle{ACM-Reference-Format}
\bibliography{sample-base}

\pagebreak

\end{document}